\documentclass{article} 
\usepackage[utf8]{inputenc} 
\usepackage{amsmath}
\usepackage{amsthm} 
\usepackage{amssymb} 
\usepackage{ setspace}
\usepackage{graphicx}
\usepackage[T1]{fontenc}

\usepackage[bitstream-charter]{mathdesign}

\usepackage[english]{babel}
\usepackage{supertabular}
\usepackage{multicol}
\usepackage{breqn}
\usepackage{xcolor}

\newcommand{\mathsym}[1]{{}}
\newcommand{\unicode}[1]{{}}

\usepackage[top=.8in, bottom=.8in, left=.8in, right=.8in]{geometry}
\usepackage{booktabs} 
\usepackage{array} 
\usepackage{paralist} 
\usepackage{verbatim} 
\usepackage{subfig} 

\usepackage{fancyhdr} 
\usepackage{sectsty}

\usepackage[nottoc,notlof,notlot]{tocbibind} 
\usepackage[titles,subfigure]{tocloft} 

\sectionfont{\large}

\title{The \textit{Skin In The Game} Heuristic for Protection Against Tail Events}
\author{Nassim N. Taleb and Constantine Sandis}
\date{October 2013}

\begin{document}
\maketitle
\setstretch{1.1}
\begin{abstract} 
\noindent
Standard economic theory makes an allowance for the agency problem, but not the compounding of moral hazard in the presence of informational opacity, particularly in what concerns high-impact events in fat tailed domains (under slow convergence for the law of large numbers). Nor did it look at exposure as a filter that removes nefarious risk takers from the system so they stop harming others. \textcolor{red}{ (In the language of probability, skin in the game creates an absorbing state for the agent, not just the principal)}. But the ancients did; so did many aspects of moral philosophy. We propose a global and morally mandatory heuristic that anyone involved in an action which can possibly generate harm for others, even probabilistically, should be required to be exposed to some damage, regardless of context. While perhaps not sufficient, the heuristic is certainly necessary hence mandatory. It is supposed to counter \textcolor{red}{voluntary and involuntary risk hiding} $-$ and risk transfer $-$ in the tails.  We link the rule to various philosophical approaches to ethics and moral luck.
\end{abstract} 

\begin{center}
 [Ethics/Epistemology/Risk Management/Probability]\\
 \textit{To Appear in Review of Behavioral Economics}, 2014, 1: 1–21
\end{center}

\bigskip

{\centering\bfseries

\par}

\section{Agency Problems and Tail Probabilities}

\begin{multicols}{2}
\noindent
The chances of informed action and prediction can be seriously increased if we better comprehend the multiple causes of ignorance. The study of ignorance, then, is of supreme importance in our individual and social lives, from health and safety measures to politics and gambling (Rescher 2009). But how are we to act in the face of all the uncertainty that remains after we have become aware of our ignorance? The idea of
\textit{skin in the game } when involving others in tail risk exposures is crucial for the well-functioning of a complex world. In an opaque system fraught with unpredictability, there is, alas, an incentive and easy opportunity for operators to hide risk: to benefit from the upside when things go well without
ever paying for the downside when one's luck runs out. 

The literature in risk, insurance, and contracts has amply dealt with the notion of information asymmetry (see Ross, 1973, Grossman and Hart, 1983, 1984, Tirole 1988,  Stiglitz 1988), but not with the consequences of deeper information opacity (in spite of getting close, as in Hölmstrom, 1979), by which tail events are impossible to figure out from watching time series and external signs: in short, in the "real world" (Taleb, 2013), the law of large numbers works very slowly, or does not work at all in the time horizon for operators, hence statistical properties involving tail events are completely opaque to the observer.  And the central problem that is missing behind the abundant research on moral hazard and information asymmetry is that these rare, unobservable events represent the bulk of the properties in some domains. We define a fat tailed domain as follows: a large share of the statistical properties come from the extremum; for a time series involving $n$ observations, as $n$ becomes large, the maximum or minimum observation will be of the same order as the sum. Excursions from the center of the distributions happen brutally and violently; the rare event dominates. And economic variables are extremely fat tailed (Mandelbrot, 1997). Further, standard economic theory makes an allowance for the agency problem, but not for the combination of agency problem, informational opacity, and fat-tailedness. It has not yet caught up that tails events are not predictable, not measurable statistically  unless one is \emph{causing} them, or involved in increasing their probability by engaging in a certain class of actions with small upside and large downside.  (Both parties may not be able to gauge probabilities in the tails of the distribution, but the agent knows which tail events do not affect him.) \textcolor{red}{Sadly, the economics literature's treatment of tail risks , or "peso problems" has been to see them as outliers to mention \textit{en passant} but  hide under the rug, or remove from analysis, rather than a core center of the modeling and decision-making, or to think in terms of robustness and sensitivity to unpredictable events.  Indeed, this pushing under the rug the determining statistical properties explains the failures of economics in mapping the real world, as witnessed by the inability of the economics establishment to see the accumulation of tail risks leading up to the financial crisis of 2008 (Taleb, 2009).} The parts of the risk and  insurance literature that have focused on tail events and extreme value theory, such as Embrechts (1997), accepts the large role of the tails, but then the users of these theories (in the applications) fall for the logical insonsistency of assuming that they can be figured out somehow: naively, since they are rare what do we know about them? The law of large numbers cannot be of help. Nor do theories have the required robustness. Alarmingly, very little has been done to make the leap that small calibration errors in models can change the probabilities (such as those involving the risks taken in Fukushima's nuclear project) from 1 in $10^6$ to 1 in 50. 

Add to the fat-tailedness the asymmetry (or skewness) of the distribution, by which a random variable can take very large values on one side, but not the other.  An operator who wants to hide risk from others can exploit skewness by creating a situation in which he has a small or bounded harm to him, and exposing others to large harm; thus exposing others to the bad side of the distributions by fooling them with the tail properties.

Finally, the economic literature focuses on incentives as encouragement or deterrent, \textcolor{red}{but not on disincentives as potent filters that remove incompetent and nefarious risk takers from the system.} Consider that the symmetry of risks incurred on the road causes the bad driver to eventually exit the system and stop killing others. An unskilled forecaster with skin-in-the-game would eventually go bankrupt or out of business.  Shielded from potentially (financially) harmful exposure, he would continue contributing to the buildup of risks in the system. \footnote{The core of the problem is as follows. There are two effects: "crooks of randomness" and "fooled of randomness" (Nicolas Tabardel, private communication). Skin in the game eliminates the first effect in the short term (standard agency problem), the second one in the long term by forcing a certain class of harmful risk takers to exit from the game.}

Hence there is no possible risk management method that can replace skin in
the game in cases where informational opacity is compounded by
informational asymmetry viz. the principal-agent problem that arises
when those who gain the upside resulting from actions performed under
some degree of uncertainty are not the same as those who incur the
downside of those same acts\footnote{Note that Pigovian mechanisms fail when, owing to opacity, the person causing the harm is not easy to identify}. For example, bankers and corporate
managers get bonuses for positive "performance", but do not have to pay
out reverse bonuses for negative performance. This gives them an
incentive to bury risks in the tails of the distribution, particularly
the left tail, thereby delaying blowups. 

The ancients were fully aware of this incentive to hide tail risks,
and implemented very simple but potent heuristics (for the
effectiveness and applicability of fast and frugal heuristics both in
general and in the moral domain, see Gigerenzer, 2010). But we find  the genesis of both moral philosophy and risk management concentrated within the same rule \footnote{Economics seems to be born out of moral philosophy (mutating into the philosophy of action via decision theory) to which was added naive and improper 19th C. statistics (Taleb, 2007, 2013). We are trying to go back to its moral philosophy roots, to which we add more sophisticated probability theory and risk management.} . About 3,800
years ago, Hammurabi{\textquoteright}s code specified that if a builder
builds a house and the house collapses and causes the death of the
owner of the house, that builder shall be put to death. This is the
best risk-management rule ever. 

What the ancients understood very well was that the
builder will always know more about the risks than the client, and can
hide sources of fragility and improve his profitability by cutting
corners. The foundation is the best place to hide such things. The
builder can also fool the inspector, for the person hiding risk has a
large informational advantage over the one who has to find it. The same
absence of personal risk is what motivates people to only appear to be
doing good, rather than to actually do it. 

Note that Hammurabi's law is not necessarily literal: damages can be "converted" into monetary compensation. Hammurabi's law is at the origin of the \textit{lex talonis} ("eye for eye", discussed further down) which, contrary to what appears at first glance, it is not literal. \textit{Tractate Bava Kama} in the Babylonian Talmud \footnote{\textit{Tractate Bava Kama}, 84a, Jerusalem: Koren Publishers,  2013.}, builds a consensus that "eye for eye" has to be figurative: what if the perpetrator of an eye injury were blind? Would he have to be released of all obligations on grounds that the injury has already been inflicted? Wouldn't this lead him to inflict damage to other people's eyesight with total impunity? Likewise, the Quran's interpretation, equally, gives the option of the injured party to pardon or alter the punishment\footnote{Quran, \textit{Surat Al-Ma'idat}, 45: "Then, whoever proves charitable and gives up on his right for reciprocation, it will be an atonement for him." (our translation).}. This nonliteral aspect of the law solves many problems of asymmetry under specialization of labor, as the deliverer of a service is not required to have the same exposure in kind, but incur risks that are costly enough to be a disincentive.

The problems and remedies are as follows: 
\smallskip

First, consider policy makers and politicians. In a decentralized
system, say municipalities, these people are typically kept in check by
feelings of shame upon harming others with their mistakes. In a large
centralized system, the sources of error are not so visible.
Spreadsheets do not make people feel shame. The penalty of shame is a
factor that counts in favour of governments (and businesses) that are
small, local, personal, and decentralized versus ones that are large,
national or multi-national, anonymous, and centralised. When the latter
fail, everybody except the culprit ends up paying the cost, leading to
national and international measures of endebtment against future generations or
 "austerity "\footnote{\ See
McQuillan (2013) and Orr (2013);
cf. the  "many hands " problem
discussed by Thompson (1987)}.These points against
 "big government " models should not be confused with the standard libertarian argument against states securing the welfare of their citizens, but only against doing so in a centralized fashion that enables people to hide behind bureaucratic anonymity. Much better to have a communitarian municipal approach:in situations in which we cannot enforce skin-in-the game we should change the system to lower the consequences of errors.

\smallskip
 Second, we misunderstand the incentive structure of corporate
managers. Counter to public perception, corporate managers are not
entrepreneurs. They are not what one could call agents of capitalism.
Between 2000 and 2010, in the United States, the stock market lost (depending
how one measures it) up to two trillion dollars for investors, compared to
leaving their funds in cash or treasury bills. It is tempting to think
that since managers are paid on incentive, they would be incurring
losses. Not at all: there is an irrational and unethical asymmetry.
Because of the embedded option in their profession, managers received
more than four hundred billion dollars in compensation. The manager who
loses money does not return his bonus or incur a negative one\footnote{There can be situations of overconfidence by which the CEOs of companies bear a disproportionately large amount of risk, by investing in their companies, as shown by Malmendier and Tate(2008, 2009), and end up taking more risk because they have skin in the game. But it remains that CEOs have optionality, as shown by the numbers above. Further, the heuristic we propose is necessary, but may not be sufficient to reduce risk, although CEOs with a poor understanding of risk have an increased probability of personal ruin. }.The
built-in optionality in the compensation of corporate managers can only
be removed by forcing them to eat some of the losses\footnote{We define "optionality" as an option-like situation by which an agent has a convex payoff, that is, has more to gain than to lose from a random variable, and thus has a positive sensitivity to the scale of the distribution, that is, can benefit from volatility and dispersion of outcomes.}.

\smallskip
Third, there is a problem with applied and academic economists,
quantitative modellers, and policy wonks. The reason economic models do
not fit reality (fat-tailed reality) is that economists have no disincentive and are never
penalized for their errors. So long as they please the journal editors, or produce cosmetically sound "scientific" papers,
their work is fine. So we end up using models such as portfolio theory and similar
methods without any remote empirical or mathematical reason. The solution is to prevent
economists from teaching practitioners, simply because they have no mechanism to exit the system in the event of causing risks that harm others. Again this brings us to decentralization by a system where policy is decided at a local level
by smaller units and hence in no need for
economists\footnote{\ \emph{A destructive combination of false rigor and lack of skin in
the game}. The disease of formalism in the application of probability to
real life by people who are not harmed by their mistakes can be
illustrated as follows, with a very sad case study. One of the most
{\textquotedbl}cited{\textquotedbl} documents in risk and quantitative
methods about {\textquotedbl}coherent measures of risk{\textquotedbl}
set strong principles on how to compute the {\textquotedbl}value at
risk{\textquotedbl} and other methods. Initially circulating in 1997,
\ the measures of tail risk -while coherent {}-have proven to be
underestimating risk at least 500 million times (sic, the number is not a typo). \ We have had a
few blowups since, including Long Term Capital Management; and we had a
few blowups before, but departments of mathematical probability were
not informed of them. As we are writing these lines, it was announced
that J.-P. Morgan made a loss that should have happened every ten
billion years. The firms employing these {\textquotedbl}risk
minds{\textquotedbl} behind the {\textquotedbl}seminal{\textquotedbl}
paper blew up and ended up bailed out by the taxpayers. But we now know
about a {\textquotedbl}coherent measure of risk{\textquotedbl}. \ This
would be the equivalent of risk managing an airplane flight by spending
resources making sure the pilot uses proper grammar when communicating
with the flight attendants, in order to {\textquotedbl}prevent
incoherence{\textquotedbl}. Clearly the problem is that tail events are
very opaque computationally, and that such misplaced precision leads to
confusion. The {\textquotedbl}seminal{\textquotedbl} paper: Artzner,
P., Delbaen, F., Eber, J. M., \& Heath, D. (1999). Coherent measures of
risk. \textit{Mathematical finance}, 9(3), 203-228.}.

\smallskip
Fourth, the predictors. Predictions in socioeconomic domains
don't work. Predictors are rarely harmed by their
predictions. Yet we know that people take more risks after they see a
numerical prediction. The solution is to ask ---and only take into
account--- what the predictor has done (what he has in his portfolio),
or is committed to doing in the future. It is unethical to drag people
into exposures without incurring losses. Further, predictors work with binary variables (Taleb and Tetlock, 2013), that is, "true" or "false" and play with the general public misunderstanding of tail events. They have the incentives to be right more often than wrong, whereas people who have skin in the game do not mind being wrong more often than they are right, provided the wins are large enough. In other words, predictors have an incentive to play the skewness game (more on the problem in section 2). The simple solution is as follows: predictors should be exposed to the variables they are predicting and should be subjected to the dictum "do not tell people what you think, tell them what you have in your portfolio" (Taleb, 2012, p.386) . 
Clearly predictions are harmful to people as, by the psychological mechanism of anchoring, they increases risk taking. 

\smallskip
Fifth, to deal with warmongers, Ralph Nader has rightly proposed that those who vote in favor of war should subject themselves (or their own kin) to the draft. 
\smallskip

We believe \textit{Skin in the game} is a heuristic for a safe and
just society. It is even more necessary under fat tailed environments. Opposed to this is the unethical practice of taking all
the praise and benefits of good fortune whilst disassociating oneself
from the results of bad luck or miscalculation. We situate our view
within the framework of ethical debates relating to the moral
significance of actions whose effects result from ignorance and \ luck.
We shall demonstrate how the idea of skin in the game can effectively
resolve debates about (a) moral luck and (b) egoism vs. altruism, while
successfully bypassing (c) debates between subjectivist and objectivist
norms of action under uncertainty, by showing how their concerns are
of no pragmatic concern.

\smallskip
\noindent
\textbf{Reputational Costs in Opaque Systems:} Note that our analysis includes costs of reputation as skin in the
game, with future earnings lowered as the result of a mistake, as with
surgeons and people subjected to visible malpractice and have to live
with the consequences. So our concern is situations in which cost
hiding is effective over and above potential costs of reputation,
either because the gains are too large with respect to these costs, or
because these reputation costs can be
{\textquotedbl}arbitraged{\textquotedbl}, by shifting blame or escaping
it altogether, because harm is not directly visible. The latter
category includes bureaucrats in non-repeat environments where the
delayed harm is not directly attributable to them. Note that in many domains the payoff can be large enough to offset reputational costs, or, as in finance and government, reputations do not seem to be aligned with effective track record.  (To use an evolutionary argument, we need to avoid a system in which those who make mistakes stay in the gene pool, but throw others out of it.)

\smallskip
\noindent
\textbf{Application of The Heuristic:} The heuristic implies that one should be the first consumer of one's product, a cook should test his own food, helicopter repairpersons should be ready to take random flights on the rotorcraft that they maintain, hedge fund managers should be maximally invested in their funds. But it does not naively imply that one should always be using one's product: a barber cannot cut his own hair, the maker of a cancer drug should not be a user of his product unless he is ill. So one should use one's products \textit{conditionally} on being called to use them. However the rule is far more rigid in matters entailing sytemic risks: simply some decisions should never be taken by a certain class of people.

\smallskip
\noindent
\textbf{Heuristic vs Regulation:} A heuristic, unlike a regulation, does not require state intervention for implementation. It is simple contract between willing individuals: "I buy your goods if you use them", or "I will  listen to your forecast if you are exposed to losses if you are wrong" and would not require the legal system any more than simple commercial transaction. It is bottom-up.
(The ancients and more-or-less ancients effectively understood the contingency and probabilistic aspect in contract law, and asymmetry under opacity, as reflected in the works of  Pierre de Jean Olivi. Also note that the foundation of maritime law has resided in skin-the-game unconditional sharing of losses, even as far in the past as 800 B.C. with the \textit{Lex Rhodia}, which stipulates that all parties involved in a transaction have skin in the game and share losses in the event of damage. The rule dates back to the Phoenician commerce and caravan trades among Semitic people. The idea is still present in Islamic finance commercial law, see Wardé, 2010 .)

\smallskip
The rest of this essay is organized as follows. First we present the epistemological dimension of the hidden payoff, expressed using the mathematics of probability, showing the gravity of the problem of hidden consequences. We present the historical background in the various philosophical branches dealing with moral luck and ethics of risk.  We conclude with the notion of heuristic as simple "convex" rule, simple in its application.

\end{multicols}

\hrule

\begin{figure}[h!]
 \centering
\includegraphics[width=.7\textwidth]{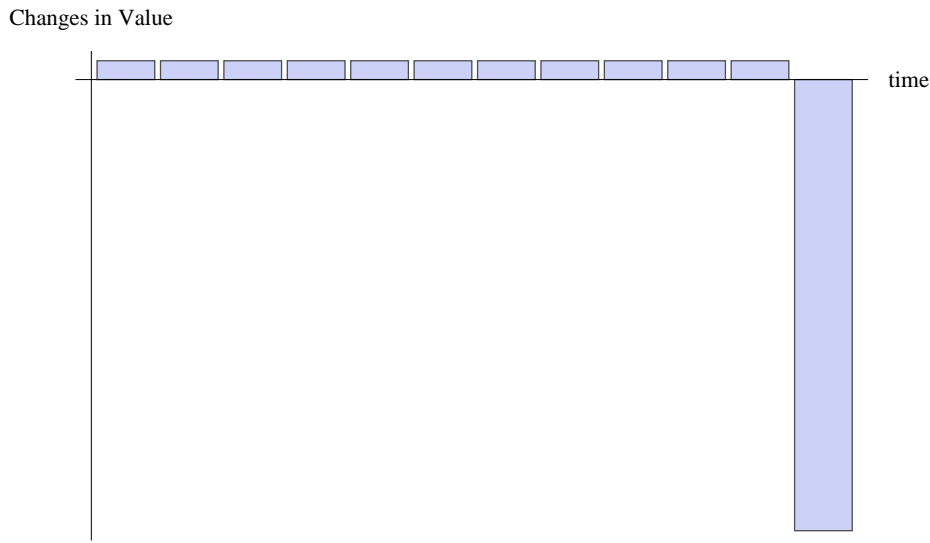}

\caption{The most effective way to maximize the expected payoff to the agent at the expense of the principal.}
\end{figure}
\noindent


\begin{multicols}{2}

\bigskip
\section{Payoff Skewness and Lack of Skin-in-the-Game}
\bigskip


This section will analyze the probabilistic mismatch or tail risks and returns in the presence of a principal-agent problem. 
\bigskip

\noindent\pmb{Transfer of Harm}: \emph{If an agent has the upside of the payoff of the random variable, with no downside, and is
judged solely on the basis of past performance, then the incentive is to hide risks in the left tail using a negatively skewed (or more generally, asymmetric) distribution for the performance. This can be generalized to any payoff for which
one does not bear the full risks and negative consequences of one's actions.}

Let { }\(P(K,M)\) be the payoff for the operator over \textit{ M} incentive periods

\begin{dmath}\label{payoff} 
P(K,M)\equiv \gamma  \sum _{i=1}^M q_{t+(i-1)\text{$\Delta $t} } \left(x_{t+i \Delta t}^j-K\right){}^+ \mathbf{1}_{\text{$\Delta $t} (i-1)+t<\tau } 
\end{dmath}
\bigskip

\noindent 
with \(X^j= (x_{t+i \Delta t}^j)_{i=1}^M \in \mathbb{R}\), i.i.d. random variables representing the distribution
of profits over a certain period \([t,t+i \text{$\Delta $t}]\), \textit{ i} $\in $ $\mathbb{N}$, \textit{ $\Delta $t} $\in $ \(\mathbb{R}^+\) and
K is a {``}hurdle{''}, $\tau $= inf\(\left\{s:\left(\sum _{z\leq s}x_z\right)<x_{\min } \right\}\) is an indicator of stopping time when past performance
conditions are not satisfied (namely, the condition of having a certain performance in a certain number of the previous years, otherwise the stream
of payoffs terminates, the game ends and the number of positive incentives stops). The constant $\gamma $ $\in $(0,1) is an {``}agent payoff{''},
or compensation rate from the performance, which does not have to be monetary (as long as it can be quantified as {``}benefit{''}). The quantity
\(q_{t+(i-1) \text{$\Delta $t}}\)\textit{  $\in $} [1,$\infty $) indicates the size of the exposure at times \textit{ t+(i-}1\textit{ ) $\Delta $t
}(because of an Ito lag, as the performance at period \textit{ s} is determined by \textit{ q} at a a strictly earlier period $<$\textit{ s})

Let \(\left\{f_j\right\}\) be the family of probability measures \(f_j\) of \(X^j\)\textit{ ,} \textit{  j} $\in $ $\mathbb{N}$. Each measure corresponds
{ }to certain mean/skewness characteristics, and we can split their properties in half on both sides of a {``}centrality{''} parameter \(K\), as
the {``}upper{''} and {``}lower{''} distributions. With some inconsequential abuse of notation we write \(dF_{j }(x)\) as \(f_j(x)\,\mathrm{d}x\), so { }\(F_j^+\)=\(\int_K^{\infty
} f_j(x) \, \,\mathrm{d}x\) and \(F_j^-\)=\(\int_{-\infty }^K f_j(x) \, \,\mathrm{d}x\) , the {``}upper{''} and {``}lower{''} distributions, each corresponding to certain
conditional expectation \(\mathbb{E}_j^+\equiv \frac{\int _K^{\infty }x f_j(x)\,\mathrm{d}x}{\int_K^{\infty } f_j(x) \, \,\mathrm{d}x}\) and { }\(\mathbb{E}_j^-\equiv
\frac{\int _{-\infty }^Kx\text{  }f_j(x)\,\mathrm{d}x}{\int_{-\infty }^K f_j(x) \, \,\mathrm{d}x}\). { }

Now define $\nu $ $\in $ \(\mathbb{R}^+\)as a K-centered nonparametric measure of asymmetry, \(\nu _j\equiv \frac{F_j^-}{F_j^+}\), with values $>$1
for positive asymmetry, and $<$1 for negative ones. Intuitively, skewness has probabilities and expectations moving in opposite directions: the larger
the negative payoff, the smaller the probability to compensate.

We do not assume a {``}fair game{''}, that is, with\textit{  }unbounded returns \textit{ m} \textit{ $\in $} (-$\infty $,$\infty $), \(F_j^+\ \mathbb{E}_j^++ F_j^-\ \mathbb{E}_j^-= m\), which we can write as  

 \[m^++m^-= m . \]

\subsubsection*{ Simple assumptions of constant \textit{ q} and simple-condition stopping time}

 Assume \textit{ q} constant, \textit{ q} =1 and simplify the stopping time condition as having no loss larger than $-K$ in the previous periods, $\tau $ =inf$\{$\((t
+ i \text{$\Delta $t}))\): \(x_{\text{$\Delta $t} (i-1)+t}<K\)$\}$, which leads to

\begin{equation}
\mathbb{E}(P(K,M))=\gamma \ \mathbb{E}_j^+  \times \mathbb{E}\left(\sum _{i=1}^M \mathbf{1}_{t+i \Delta t <\tau }\right)
\end{equation}

Since assuming independent and identically distributed agent{'}s payoffs, the expectation at stopping time corresponds to the expectation of stopping
time multiplied by the expected compensation to the agent $\gamma $ \(\mathbb{E}_j{}^+\). And \(\mathbb{E}\left(\sum _{i=1}^M \mathbf{1}_{\text{$\Delta $t} (i-1)+t<\tau }\right)= \mathbb{E}\left(\left(\sum _{i=1}^M \mathbf{1}_{\text{$\Delta $t} (i-1)+t<\tau }\right)\wedge M\right)\).

The expectation of stopping time can be written as the probability of success under the condition of no previous loss:

\[\mathbb{E}\left(\sum _{i=1}^M \ \mathbf{1}_{t+i \Delta t <\tau }\right)=\sum _{i=1}^M  F_j^+\ \mathbb{E}(\mathbf{1}_{x_{\Delta t (i-1)+t}>K}) . \]

We can express the stopping time condition in terms of uninterrupted success runs. Let $\sum $ be the ordered set of consecutive success runs \(\sum\equiv \{\{F\},\{\text{SF}\}, \{\text{SSF}\},\text{...},\{(M-1) \text{ consecutive } S,F\}\}\), where \textit{ S} is success and \textit{ F} is failure
over period \(\text{$\Delta $t}\), with associated corresponding probabilities $\{$(\(1-F_j^+\)), \(F_j^+\left(1-F_j^+\right),F_j^+{}^2\left(1-F_j^+\right),\text{...},F_j^+{}^{
M-1}\left(1-F_j^+\right)\)$\}$ ,

\begin{equation}
\sum _{i=1}^M F_j^+{}^{(i-1)}\left(1-F_j^+\right)= 1-F_j^+{}^M\simeq 1\
\end{equation}

For M large, since \(F_j^+\)\textit{  $\in $} (0,1) we can treat the previous as almost an equality,  hence:

\begin{equation*}
\mathbb{E}\left(\sum _{i=1}^M \mathbf{1}_{t+(i-1)\text{$\Delta $t} <\tau }\right)=\sum _{i=1}^M (i-1)\ F_j^+{}^{(i-1)}\left(1-F_j^+\right) \simeq\frac{F_j^+}{1-F_j^+} .
\end{equation*}

Finally, the expected payoff for the agent:

$$\mathbb{E}\left(P(K,M)\right)\simeq\text{  }\gamma\  \mathbb{E}_j^+ \frac{F_j^+}{1-F_j^+} ,$$

\noindent
which increases by i) increasing \(\mathbb{E}_j^+\) , ii) minimizing the probability of the loss \(F_j^-\), but, and that{'}s the core point, even
if i) and ii) { }take place at the expense of \textit{ m} the total expectation from the package.

Alarmingly, since \(\mathbb{E}_j^+= \frac{ m- m^-}{F_j^+}\), the agent doesn{'}t care about a degradation of the total expected return \textit{ m}
if it comes from the left side of the distribution, \(m^-\). Seen in skewness space, the expected agent payoff maximizes under the distribution \textit{
j }with the lowest value of { }\(\nu _j\) (maximal negative asymmetry). { }The total expectation of the positive-incentive without-skin-in-the-game
depends on negative skewness, not on \textit{ m}. 

\end{multicols}
\noindent
\hrule
\begin{figure}[h!]
\includegraphics{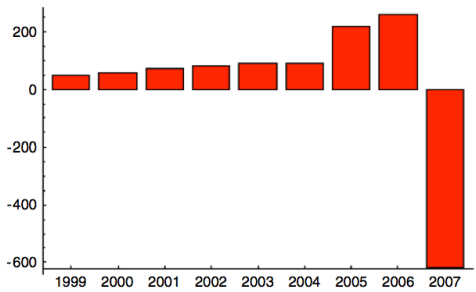}

\caption{Indy Mac, a failed firm during the subprime crisis (from Taleb 2009). It is a representative of risks that keep increasing
in the absence of losses, until the explosive blowup.}

\end{figure}
\noindent
\hrule
\begin{multicols}{2}

\subsubsection*{Multiplicative \textit{ q} and the explosivity of blowups}

Now, if there is a positive correlation between \(q\)
and past performance, or survival length, then the effect becomes multiplicative. The negative payoff becomes explosive if the allocation \textit{ q }increases with visible profitability, as seen in Figure 2 with the story of IndyMac, whose risk kept growing until the blowup\footnote{ The following sad anecdote illustrate the problem with banks. It was announces that "JPMorgan Joins BofA With Perfect Trading Record in Quarter" ( Dawn Kopecki and Hugh Son - Bloomberg News, May 9, 2013). Yet banks while "steady earners" go through long profitable periods followed by blowups; they end up losing back all cumulative profits in short episodes, just in 2008 they lost around 4.7 trillion U.S. dollars before government bailouts. The same took place in 1982-1983 and in the Savings and Loans crisis of 1991,  see Taleb (2009).}. Consider that "successful" people get more attention, more funds, more promotion. Having "beaten the odds" imparts a certain credibility. In finance we often see fund managers experience a geometric explosion of funds under management after perceived "steady" returns. Forecasters with steady strings of successes become gods. And companies that have hidden risks tend to outperform others in small samples, their executives see higher compensation. So in place of a constant exposure $q$, consider a variable one:

$$q_{\text{$\Delta $t} (i-1)+t}=q \ \omega (i) , $$

\noindent
where \(\omega (i)\) is a multiplier that increases with time, and of course naturally collapses upon blowup.

Equation \ref{payoff} becomes:
\begin{equation}
P(K,M) \equiv \gamma \text{  }\sum _{i=1}^M q\ \omega (i) \left(x_{t+i}^j{}_{\text{$\Delta $t} }-K\right){}^+\mathbf{1}_{t+(i-1)\text{$\Delta $t}
<\tau } \, ,
\end{equation}

\noindent
and the expectation, assuming the numbers of periods, \(M\) is large enough

\begin{equation}
\label{multiplier}\mathbb{E}(P(K,M))=\gamma  \  \mathbb{E}_j^+ \ q \  \mathbb{E}\left(\sum _{i=1}^M \omega (i) \  \mathbf{1}_{\text{$\Delta $t}(i-1)+t<\tau }\right) .
\end{equation}

Assuming the rate of conditional growth is a constant \(r\) $\in $ [0,$\infty $) , and making the replacement $\omega $(i)$\equiv $ \(e^{r i}\),
we can call the last term in equation  \ref{multiplier} the multiplier of the expected return to the agent:

\begin{dmath}
 \mathbb{E}\left(\sum _{i=1}^M e^{i r} \mathbf{1}_{\text{$\Delta $t} (i-1)+t<\tau }\right)=\sum _{i=1}^M (i-1)\  F_j{}^+ e^{i r} \mathbb{E}(\mathbf{1}_{x_{\Delta t (i-1)+t}>K})\ 
\end{dmath}
\begin{dmath}
$$=
\frac{ \left(F^+-1\right)\left(\left(F^+\right)^M \left(M e^{(M+1) r}-F^+ (M-1) e^{(M+2) r}\right)-F^+ e^{2 r}\right)}{\left(F^+ e^r-1\right)^2}$$
\end{dmath}


We can get the table of sensitivities for the "multiplier" of the payoff:

\begin{center}
\begin{tabular}{c|c|c|c|c}
  & \text{F=.6} & 0.7 & 0.8 & 0.9 \\
\hline
 \text{r=0} &  1.5 & 2.32 & 3.72 & 5.47 \\
 0.1 &  2.57 & 4.8 & 10.07 & 19.59 \\
 0.2 &  4.93 & 12.05 & 34.55 & 86.53 \\
 0.3 & 11.09 & 38.15 & 147.57 & 445.59 \\
\end{tabular}
\end{center}
{\centering
Table 1 Multiplicative effect of skewness
\par}

\subsubsection*{Explaining why Skewed Distributions Conceal the Mean}
Note that skewed distributions conceal their mean quite well, with \(P(X < \mathbb{E}(x))<\frac{1}{2}\) in the presence of negative skewness. And such effect increases with fat-tailedness.
Consider a negatively skewed power law distribution, say the mirror image of a standard Pareto distribution,  with maximum value \(x_{\min }\), and domain \(\left(-\infty ,x_{\min }\right]\),  with exceedance probability \(P(X > x)=\text{  }-x^{-\alpha } x_{\min }^{\alpha }\), and mean  \(-\frac{\alpha  x_{\min }}{\alpha -1}\), with \({\alpha }>1\), have a proportion  of \(1-\frac{\alpha -1}{\alpha }\) of its realizations rosier than the true mean. Note that fat-tailedness increases at lower values of  \({\alpha}\). The popular "eighty-twenty", with tail exponent \({\alpha }= 1.15\), has > 90 percent of observations above the true mean\footnote{This discussion of a warped probabilistic incentive corresponds to what John Kay has called the "Taleb distribution",  John Kay "A strategy for hedge funds and dangerous drivers", Financial Times, 16 January 2003.}.
Likewise, to consider a thinner tailed skewed distribution, for a Lognormal distribution with domain \( (-\infty, 0)\), with mean \(m=-e^{\mu +\frac{\sigma ^2}{2}}\), the probability of exceeding the mean is \(P(X>m= \frac{1}{2} \text{erfc}\left(-\frac{\sigma }{2 \sqrt{2}}\right)\), which for $\sigma=1$ is at 69\%, and for $\sigma=2$ is at 84\%.

\subsubsection*{Forecasters} 
We can see how forecasters who do not have skin in the game have the incentive of betting on the low-impact high probability event, and
ignoring the lower probability ones, even if these are high impact. There is a confusion between {``}digital payoffs{''} \(\int f_j(x) \, \,\mathrm{d}x\) and full distribution, called {``}vanilla
payoffs{''}, \(\int x f_j(x)\,\mathrm{d}x\), see Taleb and Tetlock (2013)\footnote{Money managers do not have enough skin in the game unless they are so heavily invested in their funds that they can end up in a net negative form the event. The problem is that they are judged on frequency, not payoff, and tend to cluster together in packs to mitigate losses by making them look like "industry event". Many fund managers beat the odds by selling tails, say covered writes, by which one can increase the probability of gains but possibly lower the expectation. They also have the optionality of multi-time series; they can manage to hide losing funds in the event of failure. Many fund companies bury hundreds of losing funds away, in the "cemetery of history" (Taleb, 2007) .}.

\section{Symmetrical Constraints in Moral Philosophy}
We now turn to a philosophical approach to the problem. 

The skin in the game heuristic is best viewed as a rule of thumb that
places a pragmatic \textit{constraint }on normative theories. Whatever
the best moral theory \ (consequentialism, deontology, contractualism,
virtue ethics, particularism etc.) or political ideology (socialism,
capitalism, libertarianism) might be, the
 "rule " tells us that we should be
suspicious of people who appeal to it to justify actions that pass the
cost of any risk-taking to another party whilst keeping the benefits
for themselves. At the heart of this heuristic lies a simple moral objection to
negative asymmetry that lies at the heart of some of the oldest and
most famous moral ideas, as illustrated in Table 2.
\end{multicols}

\
\begin{flushleft}
\tablehead{}
\begin{tabular}{|m{1.in}|m{1. in}|m{1. in}|m{1.in}|m{1.in}|}
\hline
~

\textbf{1} \textit{Lex Talioni}s:  "An eye
for an eye, a tooth for a tooth "\par

\centering (Exodus 21.24)\par

~
 &
~

 \textbf{2} \textit{15th Law of Holiness and Justice:}
 "Love your neighbor as yourself "\par

\centering (Leviticus 19.18) &
~

\textbf{3} \textit{Silver Rule}:  "Do not
do unto others what you would not have them do unto
you "\par

\centering (Isocrates and Hillel the Elder)\footnotemark{} &
~

 \textbf{4} \textit{Golden Rule}: Do unto others as you
would have them do unto you "\par

\centering (Matthew 7:12)\par

~
 &
~

 \textbf{5} \textit{Formula of the Universal
Law}:  "act only in accordance with that maxim through
which you can at the same time will that it become a
universal " law. " \ (Kant 1785: 4:421)
\\\hline
\end{tabular}
\end{flushleft}
\footnotetext{\ Isocrates, who was the first to express a symmetrical
principle (5th Century BC), does so in relation to emotion:
"do not do to others that which angers you when they do it to you " (Isoc3.61).}
{\centering
Table 2. Moral Symmetry
\par}
\ \

\hrule

\bigskip

\begin{multicols}{2}
\noindent
Of course the clearest examples of any rule are likely to stem from a
deontological approach, but the skin in the game constraint is not
committed to deontology. Indeed, moral symmetry is one of the key ideas
behind many forms of social contract theory (e.g.  "I
scratch your back, you scratch mine "), and different
emphases on symmetry may also be found in consequentialism (which
places the overall good above that of the agent) and virtue ethics
(which looks for an ethical mean between excess and deficiency).

\ \ \  As worded, all of the principles in the table above are
problematic. \ Take, for example, \ the fourth principle of reciprocity
in Fig.1 above. This  "golden rule "
seems to suggest that if I would like you to come up and kiss me then I
should go up to you and kiss you (regardless of whether \textit{you}
would like this). But while the precise principles may be faulted,
\ the spirit of symmetry behind them (and arguably every moral
tradition)\footnote{\ See Blackburn (2001:101).} contains much insight.
Indeed, the very plausibility of Derek Parfit "s recent
attempt to demonstrate that the best versions of the most popular
normative theories converge (Parfit 2012), must ultimately hang upon a
common spirit of this kind. As we shall we shall see, however, there
can be positive asymmetries in our behaviour, as well as negative ones.

\bigskip

\section{Altruism vs. Egoism}

Psychological Egoists claim that we always do what we most desire
(Mandeville: 1714). Those who believe in the possibility of altruism
tend do either deny this (Nagel 1970) or to distinguish between
self-centred desires and the desire to benefit others (Butler 1726). So
while it is not false to think that whether or not we ever act
altruistically is an empirical question (Slote 1964), its answer will
partly depend upon a priori distinctions between \ notions such as
those of  "desire ",
 "motivation ",
 "reason ", \ and so on. It is such
distinctions, rather than experimental research, which allow us to
recognise that while anyone who is not a sociopath will feel
contentment in helping others, it would be perverse to help others
\textit{in order} to acquire this feeling (Sandis 2012:75; cf. Broad
1930). 

\ \ \ The most pragmatic way of distinguishing between egoists and
altruists is to ask whether someone has ever voluntarily (a) paid a
cost for someone else "s benefit or (b) been willing to
reap the rewards of risk while passing the cost to another. The first,
altruistic, action is one where the agent has skin in another
person "s game (Taleb 2013), including the lives of
future generations. The second, egoistic act, is one where the person
has no skin in the game \footnote{Such altruism includes cases in which one voluntarily removes oneself from the social pool (e.g. through suicide or self-imposed exile) so as not to harm it. These should be distinguished from the agency problem in evolutionary theory}. People we call
 "saints " are frequently disposed \ to
act in the former way. Those who tend to act in the latter way we
typically call  "assholes ". In
reality, most of us are neither : we usually have skin in our own games
and those of our loved ones, but nobody else "s. On
occasion, however, even the most average of people is liable to either
slip up or rise to the occasion. Such moments are respectively marked
\ by negative or positive asymmetries (see Table 3)
\end{multicols}

\begin{centering}
\begin{tabular}{|c c c|}
\hline
\\
\textbf{No skin in the game}  & \textbf{Skin in the game } & \textbf{Skin in someone else's} 
\\
&  & \textbf{game}\\\specialrule{3pt}{1pt}{1pt}\\
\\
Selfish/egoistic & Neither egoistic nor altruistic & Selfless/altruistic \\
\\
Negative asymmetry & Symmetry (neutral) & Positive asymmetry \\
\\
Individualistic Morality & Conventional morality & Other-based morality\\
\hline
\end{tabular}

\begin{center}
Table 3: Egoism vs Altruism
\end{center}
\end{centering}

\ \
\hrule

\begin{multicols}{2}
\noindent
The middle column in the table is the largest because most of the
actions of the average person tend to fall within it. It is no wonder,
then, that the  "eye for an eye "
reciprocity it epitomises is - for better or worse - a conventional
morality. To its left lies the sort of individualistic morality
frequently associated with Nietzsche but most clearly ascribable to the
 "rational " normative egoism of Ayn
Rand (1964) and many others who maintain that  "greed
is good ". To its right lies the morality of
self-sacrifice. This comes in all sorts of stripes: Christian,
socialist, utilitarian, and so on. Needless to say, these divisions are
never as sharp in practice as they are in theory.
Rand's egoistic heroes, for example, subscribe to the
symmetrical thought that one should never demand that others take a
risk one wouldn't take oneself. Conversely, most
welfare states are run by bureaucrats with no skin in the game. Both
sides are fooling themselves.

\ \ \ The symmetrical constraint entails that we act wrongly when we
open ourselves to great harm that could have reasonably been foreseen
and avoided, but the wrongness isn "t a moral one. We
act immorally when we open \textit{others }to great risk but are only
willing to be considered as responsible for our actions if the risk
turns out not to harm anyone. Such actions involve the malignant
transfer of fragility and anti-fragility from one party to another with
the aim of getting any possible benefits of our actions without being
liable for any possible harms (Taleb 2012). This agency problem is
that of a negative asymmetry. 

\ \ \ Those who are responsible for such transfers (most predictive
analysts, economists, bankers, bureaucrats, consultants, editors,
politicians, risk vendors, and sophists) attempt to justify their
hypocrisy by appealing to bad luck and uncertainty. They offer excuses
of the  "we acted on information we believed was
correct at the time " or  "obviously it
fell way short of expectations " variety, but refuse to
accept any liability for their actions and protest wildly at the mere
thought that they should pay the cost. These may be contrasted with those who
have skin in the game viz. those who take risks for themselves and keep
their downside. Typical examples are activists, artisans, citizens (as
opposed to  "\textit{idiotes} "),
entrepreneurs, traders, and writers. The greatest contrast, however, is
with those who put their own skin in the game for the sake of others.
We call such people heroes and saints but they include not only knights
and warriors but also some maverick artists, journalists, scientists,
and writers who put their livelihood reputations on the line for the
sake of others (Taleb 2012). This all \ brings us to the so-called
 "problem of moral luck ".

\section{ Moral Luck}
Consider the case of two equally reckless drivers, only one of which
kills a pedestrian. According to Bernard Williams the unlucky driver is
morally guilty of something worse than the other driver (namely
manslaughter). Kantians, by contrast, maintain that both drivers would
only be liable for reckless driving. Both views are
confused. What we should say is that from the moral point of
view, a certain kind of reckless driving is as bad as manslaughter.
When a person drives recklessly he takes upon himself the risk of
manslaughter and is accordingly responsible for it if it happens, and
for opening himself up to it (which is just as bad from a purely
ethical point of view) if it doesn't (see Sandis
2010). Hegel got it right, then when he wrote not only that
 "[t]he laurels of mere willing are dry leaves that
never were green " but also:

\bigskip
\small
\begin{quote} It happens of course that circumstances may make an action miscarry to a
greater or lesser degree. In the case of arson, for instance, the fire
may not catch or alternatively it may take hold further than the
incendiary intended. In spite of this, however, we must not make this a
distinction between good and bad luck, since in acting a man must lay
his account with externality. The old proverb is correct:
 "A flung stone is the
devil "s ". To act is to expose oneself
to bad luck. Thus bad luck has a right over me and is an embodiment of my own willing. \end{quote}
\normalsize
\begin{flushright}
(Hegel, \textit{Philosophy of Right   }, 119A).\end{flushright}

We are not only responsible for the \textit{known }of our actions and their
effects but also for those that we \textit{ought }be aware of (even if
we are not). Our ignorance does not always relieve us of responsibility
for things we have done, because others can claim that, as rational
beings we \textit{should }know what we were doing even if we did
not.\footnote{\ For a related point see Thompson (1983).} Such is the
knowledge involved in putting other people "s lives at
risk with no skin (of our own) in the game.Hegel's solution famously offers two aspects of any
given act: \textit{Tat }(deed) corresponding to the objective (which I
am causally responsible for), and \textit{Handlung }(action)
corresponding to the subjective (which can be morally imputed to me);
rights relating to the latter in turn dividing into ones relating to
various elements of the self such as knowledge, intention, and purpose
(PR 115, 117, \& 120; see also 118A). 

\ \ Bad luck is no excuse when it could have been reasonably foreseen.
Foresight should not be restricted here to a particular event. If I
know that 1/1000 actions of type A will have a tragic result it is not
acceptable to perform thousands of these actions on the grounds that
for each one there is only a probability of 1/1000 that something will
go wrong. \ The greater the potential disaster the smaller the
probability has to be for an act that could bring it about to be
immoral. There is an inverse symmetry between the acceptable
probability of risk and the weight of the potential damage being
assessed. 

\ \ All action is, to varying degrees, \textit{exposition }luck and must
be judged accordingly. When we take a risk we cannot wash our hands of
the consequences on others and hide behind masks of expectation,
intention, ignorance, luck, uncertainty, and so on.
The central point bears repeating here: asymmetry in
taking risks without having skin in the game is an unethical one. Any
system deemed  "too big too fail " not
only encourages but demands that we live according to such skinless
asymmetry. The real black swan event of the 21st century is not that
any financial crisis occurred (which was predictable) but that there
was no full-blown revolution against the governments which continue to
encourage  "idiotes " to gamble with
other people "s lives and money.

\section {
 Objectivism vs. Subjectivism}

The ethics of risk is frequently thought of as a branch of moral
philosophy concerned with abstract principles that tell us how we ought
to act when we lack (or do not know whether or not we lack) information
that is relevant to our choice (e.g. Altham 1984 who makes a technical
distinction between mere risk and general uncertainty). Far from being
infrequent, such scenarios are the norm and can only be excluded in
controlled thought experiments. In an important sense, then, all acts
are performed under uncertainty, which is not to say that we never know
what the consequences of our actions will be (see 2002: 233). This
raises the problem of how we ought to act in the face of known
ignorance. \ The skin in the game ethic bypasses the issue, revealing
it to be pragmatically irrelevant.

\ \ \  The worry is that of whether a person{\textquoteleft}s obligation
to perform (or omit from performing) some action depends
 "on certain characteristics of the situation in which
he is, or on certain characteristics of his thought about the
situation " (Prichard 1932: 84).Objectivists (such as
Sidgwick and Parfit) claim that we ought to do whatever is \textit{in
fact} be best, even when we cannot be reasonably expected to know what
this is. \ By contrast subjectivists (including Ross 1939 and Prichard
1932) claim that we ought to do whatever we \textit{believe} will be
best. 

\ \ \ The difficulty of choosing between these positions is supposed to
stem from two considerations that are in tension. On the one hand, we
want to leave room for the thought that we can be \textit{\ }what we
ought to do. The fact that what we \textit{believe }ought to do and
what we \textit{actually }to do can come apart in this way \ seem to
lend credence to objectivism. On the other hand, there is the
procedural obstacle of the impossibility of stepping out of
one "s own mind in order to compare reality with
one "s impressions of it. Thus the objective view
\ appears to entail the absurd view that  "although we
may have duties, we cannot know but can only believe that we have; and
therefore we are rendered uncertain whether we, or anyone else, has
ever had, or will ever have a duty " (Prichard
1932:88-9).\footnote{\ Ross 1930 rightly (but for the wrong reason)
suggests that objectivists and subjectivists are talking at cross
purposes. Cf. Zimmerman (2008: 1-2).} A parallel absurdity is implied
in this rhetorical question posed by Jonathan Dancy:
 "Suppose that, unknown ...to me, someone has been
buried alive in my garden during the night. Could this make it wrong of
me to go away for a fortnight "s
holiday? " (Dancy 2000:57). Prospectivists, most
prominently Michael Zimmerman, attempt to avoid this dilemma by arguing
that we ought to perform whichever action it is most reasonable to
\textit{expect} will be the best. 

\ \  Such academic debates have little pragmatic weight. All three views
share the common mistaken assumption that they are each motivated by
the same notions of  "what one ought to
do " when there are actually three different concepts
at play:

\bigskip

\ (i) \textbf{Objectivists} equate \textit{what }we ought to do with
whichever action turns out to be best. \ This is what we should
\textit{aim} at when we act. 

\ (ii)\textbf{ Subjectivists} equate what we ought to do with whatever
we judge to be best. This the only way \textit{through which} we can
aim at what is best. 

\ (iii) \textbf{Prospectivists} equate what we ought to do with what we
can rationally expect to be best. This view attempts to reconcile
objectivist and subjectivist intuitions that are only in tension
because of the aforementioned assumption.

\bigskip

\ Whereas Objectivists are concerned with the rightness of the things we
do (typically thought to be universals), Prospectivists and
Subjectivists are concerned with the rightness of our acts of
\textit{doing} these things (typically thought to be particulars). Yet
it is possible that one rightly acts in doing something that results in
negative value and, by the same token, that one acts wrongly in doing
something that turns out positively.\footnote{\ This point runs
parallel to the distinction between a belief and a believ\textit{ing}
being justified (e.g. as introduced in the literature on Gettier
cases).} 

\ \ \  Given that one can do the right thing for the wrong reason, the
deontic question of \textit{what }the right thing to do is should
therefore be \ distinguished from the evaluative question of when one
is acting right\textit{ly}. The evaluative question is best answered via
an account of how and when people and institutions are \textit{liable
}for choices they make under uncertainty. \ We have sought to answer
the question (e.g. in the case of moral luck) via the skin in the game
principle. Strictly speaking, this necessary (though not sufficient)
moral heuristic is not about action but about dispositions. Indeed, it
relates directly to the virtue of being such that the system will not
only \textit{survive }uncertainty, randomness, and \textit{volatility}
but will actually benefit from it.\footnote{\ One can, of course,
render this into a principle about action ( "act in
whatever way renders you anti-fragile ") but such a
principle treats anti-fragility as the ultimate end-in-itself\textit{
}whereas it is best to treat it as a \ property whose value is derived
from its effects.} Skin in the game heuristics follow directly from the
principle of antifragility.

\bigskip

\section*{Acknowledgments}
Jon Elster, David Chambliss Johnson, Gur Huberman, Raphael Douady, Robert Shaw, Ralph Nader, Barkley Rosser, James Franklin, Russell Roberts, Marc Abrahams,Andreas Lind and Elias Korosis.

\bigskip

\section*{References}
\small
\setlength{\parindent}{0pt}
\setstretch{1.2}

Altham, J. E. J. (1984),  "Ethics of
Risk ", \textit{Proceedings of the Aristotelian
Society}, New Series, Vol. 84 (1983 - 1984),15-29.

Blackburn,S. (2001),\textit{Ethics: A Very Short
Introduction}, (Oxford: Oxford University Press).

Broad, C. D. (1930), \textit{Five Types of Ethical Theory}.
(London: Kegan Paul).

Butler, J. (1726),  " Sermons I
and XI " in his \textit{Fifteen Sermons Preached at the Rolls
Chapel} , in\textit{The Works of Bishop Butler}, ed. J. H. Bernard (London:
Macmillan).

Dancy, J. (2000), \textit{Practical Reality}(Oxford: Oxford University
Press).

\_\_\_\_\_ (2002),  "Prichard on Duty and Ignorance of
Fact " in P. Stratton-Lake (ed.), \textit{Ethical
Intuitionism: Re-evaluations}. Oxford: Oxford University Press. 

Embrechts, P. (1997). \textit{Modelling extremal events: for insurance and finance}. Springer.

Gigerenzer, G. (2010), {\textquotedbl}Moral Satisficing: Rethinking
Moral Behavior as Bounded Rationality{\textquotedbl}, \textit{Topics in
Cognitive Science 2} (2010) 528-554

Grossman, S. J., and Hart, O. D. (1983). "An analysis of the principal-agent problem". \textit{Econometrica}, 7-45.

Grossman, S. J., and Hart, O. D. (1983). "Implicit contracts under asymmetric information". \textit{The Quarterly Journal of Economics}, 123-156.

Hegel, G. W. F. (1991), \textit{Elements of the Philosophy of Right},
trans. H. B. Nisbet. Cambridge: Cambridge University Press. 

Hölmstrom, B. (1979). Moral hazard and observability. \textit{The Bell Journal of Economics}, 74-91.

Isocrates (1980), with an English Translation in three volumes, by G.
Norlin (Cambridge MA: Harvard University Press).

Kant, I, (1785), \textit{Groundwork for the Metaphysic of Morals, }. A.
Zweig, ed. T. E. Hill, Jr. \& A. Zweig.\textit{ }(Oxford: Oxford
University Press), 2002\textit{.}

Malmendier, U., and G. Tate, 2008, “Who Makes Acquisitions? CEO Overconfidence and the Market’s Reaction.” Journal of Financial Economics 89(1):
20–43.

Malmendier, U., and G. Tate, 2009, “Superstar CEOs.” Quarterly Journal of Economics 124(4): 1593–1638.

Mandelbrot, Benoit B., 1997, \textit{Fractals and Scaling in Finance: Discontinuity, Concentration, Risk}. New York: Springer-Verlag.

Mandelbrot, Benoit B., and N. N. Taleb, 2010, “Random Jump, Not Random Walk.” In Richard Herring, ed.,\textit{The Known, the Unknown, and the Unknowable}. Princeton, N.J.: Princeton University Press.

Mandeville, B. (1714), \textit{The Fable of the Bees} (London: J
Roberts).

McQuillan,M. (2013),  "The Severity of
Austerity ", \textit{Times Higher Education}, 3-9 Jan,
30-35.

Nagel, T. (1970, \textit{The Possibility of Altruism} (New Jersey:
Princeton University Press).

Nussbaum, M. (2001),\textit{The Fragility of Goodness: Luck and Ethics
in Greek Tragedy and Philosophy}, rev. edition. Cambridge: Cambridge
University Press. 

Orr,D. (2013),  "It wasn "t labour who
spent too much, it was the banks. How did we forget
this? ", \textit{The Guardian}, Sat 5 Jan, p.39.

Parfit, D. (2012), \textit{On What Matters}, Vols. I \& II (Oxford:
Oxford University Press).

Pratt, J. W., Zeckhauser, R., and Arrow, K. J. (1985). \textit{Principals and agents: The structure of business}. Harvard Business Press.

Prichard, H. A. (2002),  "Duty and Ignorance of
Fact " \ in \textit{Moral Writings}, ed. J. MacAdam.
Oxford: Oxford University Press.

Rand, A. (1964), \textit{The Virtue of Selfishness}: \textit{A New
Concept of Egoism} (New York:Signet).

Rescher, N. (2009), \textit{Ignorance: On the Wider Implications of
Deficient Knowledge} (Pittsburgh: University of Pittsburgh Press).

Ross, S. A. (1973). The economic theory of agency: The principal's problem. \textit{The American Economic Review}, 63(2), 134-139.

Ross, W. D. (1930), \textit{The Right and the Good}, rev. ed. 2002, ed.
P. Stratton-Lake. 

\_\_\_\_\_ (1939), \textit{The Foundations of Ethics}. Oxford: Clarendon
Press.

Sandis, C. (2010) , {\textquoteleft}The Man Who Mistook his Handlung
for a Tat: Hegel on Oedipus and Other Tragic Thebans{\textquoteright},
\textit{Bulletin of the Hegel Society of Great Britain}, No. 62, 35-60.

\_\_\_\_\_ (2012), \textit{The Things We Do and Why We Do Them}(Palgrave
Macmillan, 2012).

Slote, M. A. (1964),  "An
Empirical Basis for Psychological Egoism ",\textit{Journal of
Philosophy} 61, 530-537. 

Stiglitz, J. E. (1988). \textit{Principal and agent. In the New Palgrave: A dictionary of economics}, Vol 3, London: Macmillan.

Taleb, N N (2007), \textit{The Black Swan}, Penguin and Random House.

Taleb, N. N. (2009), Errors, robustness, and the fourth quadrant.\textit{ International Journal of Forecasting}, 25(4), 744-759.

Taleb, N.N. (2012), \textit{Antifragile: Things That Gain From Disorder}, Penguin and Random House.

Taleb, N.N. (2013),\textit{Probability and Risk in the Real World, vol 1: Fat Tails}.

Taleb, N.N., and Tetlock, P. E,  (2013) On the Difference between Binary Prediction and True Exposure with Implications for Forecasting Tournaments and Decision Making Research (June 25, 2013). Available at SSRN: http://ssrn.com/abstract=2284964 or http://dx.doi.org/10.2139/ssrn.2284964

Thompson, D.F. (1983),  "Ascribing Responsibility to
Advisers in Government ", \textit{Ethics }93 ( 3 ), 546
-- 60.

\_\_\_\_\_ (1987), \textit{Political Ethics and Public Office}
(Cambridge, MA: Harvard University Press).

Tirole, J. (1988). \textit{The Theory of Industrial Organization}. Cambridge: MIT press.

Wardé, I. (2010),\textit{Islamic finance in the global economy}. Edinburgh University Press.

Williams, B. (1993), \textit{Shame and Necessity}. Cambridge: Cambridge
University Press. 

Zimmerman, M.J. (2008), \textit{Living with Uncertainty: The Moral
Significance of Ignorance} (Cambridge: Cambridge University Press).

\end{multicols}

\end{document}